\documentclass[aps,eqsecnum,preprint,floats,epsf,epsfig,nofootinbib]{revtex4}
\usepackage{graphicx}% Include figure files
\usepackage{epsfig}

\begin{document}
\newskip\eqnskip
\eqnskip 0.5\baselineskip
\def\nslash{\rlap{\hspace{0.02cm}/}{n}}
\def\pslash{\rlap{\hspace{0.02cm}/}{p}}
\def\beq{\begin{eqnarray}}
\def\eeq{\end{eqnarray}}
\def\P{{\cal P}}
\def\pr{^{\prime}}
\def\GeV{{\rm GeV} }
\def\ln{{\rm ln}}
\def\ds{\displaystyle}
\def\skipback{\vskip -\parskip}
\def\non{\nonumber}
\def\dirac#1{#1\llap{/}}
\def\as{\alpha_s}
\def\pv#1{\vec{#1}_\perp}
\def\lqcd{\Lambda_{\rm QCD}}
\def\la{\langle}
\def\ra{\rangle}
\def\ln{\rm ln}
\def\t{\perp}
\def\ep{\epsilon}
\def\bi{\bibitem}

\newcommand\epjc{Eur.\ Phys.\ J.\ C }
\newcommand\jhep{J.\ High Ener.\ Phys.\ }
\newcommand\npb{Nucl.\ Phys.\ B }
\newcommand\npps{Nucl.\ Phys.\ B (Proc.\ Suppl.) }
\newcommand\plb{Phys.\ Lett.\ B }
\newcommand\zpc{Z.\ Phys.\ C }

 \centerline{\Large\bf Covariant light-front approach for heavy quarkonium: }
 \centerline{\Large\bf decay constants, $P\to \gamma\gamma$ and $V\to P\gamma$ }

\vskip 1.4cm

\centerline{\bf  Chien-Wen Hwang$^{a,b}$ \footnote{Email:
 t2732@nknucc.nknu.edu.tw } and
 Zheng-Tao Wei$^{b,c,d}$ \footnote{Email: weizt@phys.sinica.edu.tw}  }

\vskip 0.5cm

\centerline{$^a$ Department of Physics, National Kaohsiung Normal
 University, Kaohsiung 802, Taiwan}

\bigskip
\centerline{$^b$ National Center for Theoretical Sciences, National
 Cheng-Kung University, Tainan 701, Taiwan}

\bigskip
\centerline{$^c$ Institute of Physics, Academia Sinica, Taipei 115,
  Taiwan }

\bigskip
\centerline{$^d$ Department of Physics, Nankai University, Tianjin
 300071, China }

\vskip 1.2cm

\centerline{\bf Abstract}
\bigskip
\small

\noindent The light-front approach is a relativistic quark model and
offers many insights to the internal structures of the hadronic
bound states. In this study, we apply the covariant light-front
approach to ground-state heavy quarkonium. The pesudoscalar and
vector meson decay constants are discussed. We present a detailed
study of two-photon annihilation $P\to \gamma\gamma$ and magnetic
dipole transition $V\to P\gamma$ processes.  The numerical
predictions of the light-front approach are consistent with the
experimental data and those in other approaches. The relations of
the light-front approach with the other methods are discussed in
brief.

\eject

%%%%%%%%%%%%%%%%%%%%%
\section{Introduction}
%%%%%%%%%%%%%%%%%%%%%

Heavy quarkonium provides a unique laboratory to study Quantum Chromodynamics (QCD) for the bound
states of heavy quark-antiquark system. The nature that the heavy quarkonium is relevant to a
non-relativistic treatment had been known for a long time \cite{QR}. Although the non-relativistic
QCD (NRQCD), an effective field theory, is a powerful theoretical tool to separate the high energy
modes from the low energy contributions, the calculations of the low energy  hadronic matrix elements
rely on model-dependent non-perturbative methods in most cases. From the point of view of the
non-perturbative QCD, there is no one method which is uniquely superior over the others. Many methods
were employed in heavy quarkonium physics, such as lattice QCD, quark-potential model, etc. (for a
recent review see \cite{HQ}). The light-front quark model, in which a hadronic matrix element is
represented as the overlap of wave functions, offers many insights into the internal structures of
the bound states. In this study, we will explore the heavy quarkonium from a quark model on the light
front.

The light-front QCD has been developed as a promising analytic
method for solving the non-perturbative problems of hadron physics
\cite{BPP}. The aim of the light-front QCD is to describe the
hadronic bound states in terms of their fundamental quark and gluon
degrees of freedom. It may be the only possible method that the low
energy quark model and the high energy parton model can be
reconciled. For the hard processes with large momentum transferred,
the light-front QCD reduces to perturbative QCD (pQCD) which
factorize the physical quantity into a convolution of the hard
scattering kernel and the distribution amplitudes (or functions). In
general, the basic ingredient in light-front QCD is the relativistic
hadron wave functions which generalize the distribution amplitudes
(or functions) by including the transverse momentum distributions.
It contains all information of a hadron from its constituents. The
hadronic quantities are represented by the overlaps of wave
functions and can be derived in principle.

The light-front quark model is the only relativistic quark model in
which a consistent and fully relativistic treatment of quark spins
and the center-of-mass motion can be carried out \cite{LFQM}. This
model has many advantages. For example, the light-front wave
function is manifestly Lorentz invariant as it is expressed in terms
of the internal momentum fraction variables which is independent of
the total hadron momentum. Moreover, hadron spin can also be
correctly constructed using the so-called Melosh rotation. This
model had been successfully applied to calculate many
phenomenologically important meson decay constants and hadronic form
factors \cite{Jaus1, CCH1, Jaus2, CCH2, Hwang}.

On the light front, the non-relativistic nature of a heavy
quarkonium is  represented by that the light-front momentum
fractions of the quark and antiquark is close to $1/2$ and the
relative transverse and the $z-$direction momenta are much smaller
than the heavy quark mass. The Lorentz invariant light-front wave
function and the light-front formulations provide a systematic way
to include the relativistic corrections. There is no conceptual
problem to extend the light-front approach into the heavy
quarkonium. We will apply the covariant light-front approach
\cite{Jaus2, CCH2} to the ground-state $s$-wave mesons which include
$^1S_0$ pseudoscalar mesons ($P$) $\eta_c, \eta_b$ and $^3S_1$
vector mesons ($V$) $J/\psi, \Upsilon(1S)$ as our first-step study
along this direction. The main purposes of this study are threefold:
(1) Is the light-front approach applicable into the heavy
quarkonium? In concept, the light-front quark model is the
relativistic generalization of the non-relativistic quark model. The
phenomenological success of the previous non-relativistic
quark-potential model should be reproduced in the light-front
approach. In particular, we will examine the validity of the
light-front approach in three types of quantities: decay constants,
two-photon annihilation $P\to \gamma\gamma$ and magnetic dipole
transition $V\to P\gamma$. In most literatures, these processes were
explored separately \cite{KMRR, AB, BJV, DER}. To study them
simultaneously can better constrain the phenomenological parameters
and check the consistency of the theory predictions. (2) The
$\eta_b$ meson has still not been observed in experiment
\cite{ALEPH}. We will present the numerical prediction for the
branching ratios for $\eta_b\to\gamma\gamma$ and
$\Upsilon\to\eta_b\gamma$ processes. (3) What is the relation of the
light-front approach with the other approaches? In the
non-relativistic approximations, the light-front approach will be
closely related with the non-relativistic quark-potential approach.
For the process of $P\to\gamma\gamma$ which is light-front
dominated, the light-front approach reduce to the model-independent
pQCD.

The paper is organized as follows. In Sec. II, we give a detailed
presentation of the covariant light-front approach for heavy
quarkonium. It contains a brief review of the light-front framework
and the light-front analysis for the decay constants of $P$ and $V$
mesons and the processes $P\to\gamma\gamma$, $V\to P\gamma$. In Sec.
III, the relations of the light-front approach with the
non-relativistic approach and pQCD are discussed. In Sec. IV, the
numerical results and discussions are presented. Finally, the
conclusions are given in Sec. V.

%%%%%%%%%%%%%%%%%%%%%%%%%%%%%%%%%%%%%%%%%%%%%%%%%%%%%%
\section{Formalism of covariant light-front approach}
%%%%%%%%%%%%%%%%%%%%%%%%%%%%%%%%%%%%%%%%%%%%%%%%%%%%%%

\subsection{General formalism}

A heavy quarkonium is the hadronic bound state of heavy quark and
antiquark. In this system, the valence quarks have equal masses
$m_1=m_2=m$ with $m$ the mass of heavy quark $c$ or $b$. Thus the
formulae associated with the term $(m_1-m_2)$ vanish and will lead
to some simplifications. In this section, we will give the formulae
specially for the quarkonium system.

The momentum of a particle is given in terms of light-front
component by  $k=(k^-, k^+, k_\t)$ where $k^{\pm}=k^0\pm k^3$ and
$k_\t=(k^1, k^2)$, and the light-front vector is written as $\tilde
k=(k^+, k_{\t})$. The longitudinal component $k^+$ is restricted to
be positive, i.e., $k^+> 0$ for the massive particle. By this way,
the physical vacuum of light-front QCD is trivial except the zero
longitudinal momentum modes (zero-mode). We will study a meson with
total momentum $P$ and two constituents, quark and antiquark whose
momenta are $p_1$ and $p_2$, respectively. In order to describe the
internal motion of the constituents, it is crucial to introduce the
intrinsic variables $(x_i, p_{\t})$ through
 \beq
 && p_1^+=x_1 P^+,  \qquad \qquad  p_{1\t}=x_1 P_{\bot}+ p_{\t};  \non \\
 && p_2^+=x_2 P^+,  \qquad \qquad  p_{2\t}=x_2 P_{\bot}- p_{\t},
 \eeq
where $x_i$ are the light-front momentum fractions and they satisfy
$0<x_1, x_2<1$ and $x_1+x_2=1$. The invariant mass $M_0$ of the
constituents and the relative momentum in $z$ direction $p_z$ can be
written as
 \beq \label{eq:Mpz}
  M_0^2=\frac{p_{\t}^2+m^2}{x_1 x_2}, \qquad ~~~
  p_z=(x_2-\frac{1}{2})M_0.
 \eeq
The invariant mass $M_0$ of $q\bar q$ is in general different from
the mass $M$ of meson which satisfies $M^2=P^2$. This is due to the
fact that the meson, quark and antiquark can not be on-shell
simultaneously. The momenta $p_{\t}$ and $p_z$ constitute a momentum
vector $\vec p=(p_{\t}, p_z)$ which represents the relative momenta
in the transverse and $z$ directions, respectively.  The energy of
the quark and antiquark $e_1=e_2\equiv e$ can be obtained from their
relative momenta,
 \beq
 e=\sqrt{m^2+p_{\t}^2+p_z^2}.
 \eeq
It is straightforward to find that
 \beq
 x_1=\frac{e-p_z}{2e},~
x_2=\frac{e+p_z}{2e},~ e=\frac{M_0}{2}.
 \eeq

To calculate the decay constants or decay amplitude, the Feynman
rules for the vertices of quark-antiquark coupling to the meson
bound state are required. In the following formulations, we will
follow the notations in \cite{CCH2}. The vertices $\Gamma_M$ for the
incoming meson $M$ are given as
 \beq \label{eq:HM}
 H_P \gamma_5  \qquad \qquad   \qquad \qquad
 &&{\rm for~ }P; \non \\
 iH_V\Big [ \gamma_{\mu}-\frac{1}{W_V}(p_1-p_2)_{\mu} \Big ]
  \qquad &&{\rm for~ }V.
 \eeq
After performing a one-loop contour integral to be discussed below which amounts to make one quark or
antiquark on its mass-shell, the function $H_M$ and the parameter $W_V$ are reduced to $h_M$ and
$w_V$, respectively, and they are written by
 \beq \label{eq:htowf}
 h_P=h_V=(M^2-M_0^2)\sqrt{\frac{x_1 x_2}{N_c}}\frac{1}{\sqrt 2
      M_0}\phi(x_2,p_\bot),
 \eeq
and
 \beq
  w_V=M_0+2m.
 \eeq
The form of the function $h_M$ and the Feynman rule for $\Gamma_M$
are derived from the light-front wave function which describes a
meson bound state in terms of a quark $q_1$ and an antiquark $\bar
q_2$. The light-front wave function contains two parts: one is the
momentum distribution amplitude $\phi(x_2,p_\bot)$ which is the
central ingredient in light-front QCD, the other is a spin wave
function which constructs a state of definite spin ($S, S_z$) out of
light front helicity eigenstates ($\lambda_1, \lambda_2$). The spin
wave function is constructed by using the Melosh transformation and
its spin structure has been contained in Eq. (\ref{eq:HM}).
%It is the generalization of the distribution amplitude $\phi(x)$ of pQCD
%method.

The momentum distribution amplitude $\phi(x_2,p_\bot)$ is the
generalization of the distribution amplitude $\phi(x)$ of the pQCD
method and can be chosen to be normalizable, i.e., it satisfies
 \beq \label{eq:Norm1}
 \int \frac{dx d^2 p_{\bot}}{2 (2\pi)^3}
   |\phi(x,p_{\bot})|^2=1.
 \eeq
In principle, $\phi(x_2,p_\bot)$ is obtained by solving the light-front QCD bound state equations
$H_{LF}|\Psi\rangle=M|\Psi\rangle$ which is the familiar Schr\"{o}dinger equation in ordinary quantum
mechanics and $H_{LF}$ is the light-front Hamiltonian. To see the explicit form of the light-front
bound state equation, let us consider a quarkonium wave function. The light-front bound state
equation can be expressed as:
\begin{equation}
\begin{array}{l}
  \biggl ( M^2-\ds\sum_i \ds\frac{\,k_{i\bot}^2+m^2_i\,}{x_i}\biggr )
  \left (
\begin{array}{c}
  \Psi_{q\bar{q}} \\
  \Psi_{q\bar{q}g} \\
  \vdots
\end{array}
  \right ) \\[2.5\eqnskip]
  \hspace{0.8cm} = \left (
\begin{array}{ccc}
  \langle q\bar{q}|H_{int}|q\bar{q} \rangle & \langle
q\bar{q}|H_{int}|q\bar{q}g\rangle & \cdots \\
  \langle q\bar{q}g|H_{int}|q\bar{q} \rangle & \cdots & \\
  \vdots & &
\end{array}
  \right ) \left (
\begin{array}{c}
  \Psi_{q\bar{q}} \\
  \Psi_{q\bar{q}g} \\
  \vdots
\end{array}
  \right ) \, .
\end{array}
\label{exact}
\end{equation}
Of course, to exactly solve the above equation for the whole Fock space is still impossible.
Currently, two approaches have been developed. One is given by Brodsky and Pauli
\cite{PB,TBP,KP,KPP}, the so-called discretize light-front approach, the other by Perry, Harindranath
and Wilson \cite{PHW,PH,GHPSW}, based on the old idea of the Tamm-Dancoff approach \cite{T,D} that
truncates the Fock space to only include these Fock states with a small number of particles.
Furthermore, if one can eliminate all the high order Fock space sectors (approximately) by an
effective two-body interaction kernel, the light-front bound state equation is reduced to the
light-front Bethe-Salpeter equation:
\begin{equation}
  \biggl ( M^2-\ds\frac{\,k_{\bot}^2+m^2\,}{\,x(1-x)\,}\biggr )
  \Psi_{q\bar{q}}(x,k_{\bot})= \ds\int \ds\frac{\,dyd^2k'_{\bot}\,}{2(2\pi)^3}
  V_{eff}(x,k_{\bot},y,k'_{\bot})\Psi_{q\bar{q}}(y,k'_{\bot}) \, . \label{Bethe}
\end{equation}
One may solve the Bethe-Salpeter equation for finding the relativistic bound states. However, the
Bethe-Salpeter equation only provides the amplitude of a Fock sector in the bound states so that it
cannot be normalized. In other words, the Bethe-Salpeter amplitudes do not have the precise meaning
of wave functions for particles. In addition, the advantage for Eq. (\ref{exact}) with the
Tamm-Dancoff approximation is that it provides a reliable way to study the contribution of Fock
states which contain more particles step by step by increasing the size of truncated Fock space,
while the Bethe-Salpeter equation Eq. (\ref{Bethe}) lacks such an ability. Some studies on
nonperturbative features of light-front dynamics were focused on the 1+1 field theory. Typical
examples are: the discretized light-front quantization approach for the bound states in the 1+1 field
theory developed by Pauli and Brodsky \cite{PB,EPB}, the light-front Tamm-Dancoff approach for bound
state Fock space truncation discussed by Perry {\em et al}. \cite{PHW,PH}. However, at the present
time, how to solve for the bound states from 3+1 QCD is still unknown. We are satisfied with
utilizing some phenomenological momentum distribution amplitudes which have been constructed
phenomenologically in describing hadrons. One widely used form is the Gaussian-type which we will
employ in the application of covariant light-front approach.

%%%%%%%%%%%%%%%%%%%%%%%%%%%%%
\subsection{Decay constants}
\begin{figure}[htbp]
\includegraphics*[width=2.1in]{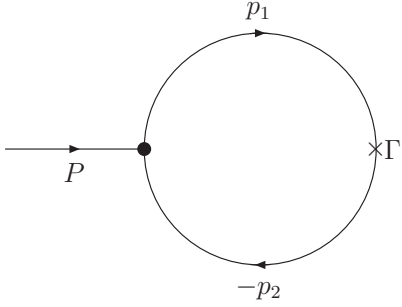}  \caption{Feynman diagram for meson
decay constants, where $P$ is the momentum of meson, $p_1$ is the
quark momentum, $p_2$ is the antiquark momentum and $\Gamma$ denotes
the corresponding $V$-$A$ current.}
 \label{fig:dc}
\end{figure}

%%%%%%%%%%%%%%%%%%%%%%%%%%%%%
In general, the decay constants of mesons $f_{P,V}$ are defined by the
matrix elements for $P$ and
$V$ mesons
 \beq
 \la 0|A_\mu |P(P) \ra &=& if_P P_\mu, \non \\
 \la 0|V_\mu |V(P) \ra &=& M_V f_V \ep_\mu.
 \eeq
where $P_\mu$ is the momentum of meson and $\ep_\mu$ is the
polarization vector of $V$ meson. The Feynman diagram which
contributes to $f_{P,V}$ is depicted in Fig. \ref{fig:dc}.
%%%%%%%%%%%%%%%%%%%%%%%%%%%%%%%%%%%%%%%%%%%%%%%%%%%%%%%%%%%%%%%%%%%%%%%%%
%%%%%%%%%%%%%%%%%%%%%%%%%%%%%%%%%%%%%%%%%%%%%%%%%%%%%%%%%%%%%%%%%%%%%%%%%
The meson decay constant plays an important role in determining the
parameters of the distribution function $\phi(x_2,p_\bot)$, in
particular, the quark mass and a parameter $\beta$ characterizing
the hadronic ``size" for a Gaussian wave function. The decay
constants have been calculated in \cite{Jaus2, CCH2} and are the
same as our results. Thus we simply provide the formulae for
$f_{P,V}$ here.
%The analysis of $P\to \gamma\gamma$ and $V\to P\gamma$ are new and
%constitutes our main objects in
%this chapter.

For a pseudoscalar quarkonium, the decay constant is represented by
 \beq \label{eq:dcP}
 f_P&=&\frac{\sqrt{2N_c}}{8\pi^3}\int dx_2 d^2 p_\t
 \frac{m}{\sqrt{x_1x_2} M_0}\phi_P(x_2,p_{\bot})\non\\
 &=&\frac{\sqrt{2N_c}}{8\pi^3}\int dx_2 d^2 p_\t
 \frac{m}{\sqrt{m^2+p_\t^2}}\phi_P(x_2,p_{\bot}).
 \eeq
where $N_c=3$ is the color number and $m$ denotes the mass of the heavy quark. In Eq. (\ref{eq:dcP}),
we have used the relation
 \beq
 M_0\sqrt{x_1x_2}=\sqrt{m^2+p_\t^2}
 \eeq
for a quarkonium .

For the vector meson, the decay constant in the covariant approach
is represented by
 \beq \label{eq:dcV}
 f_V=\frac{\sqrt{2N_c}}{8\pi^3 M}\int dx_2 d^2 p_\t
     \frac{1}{\sqrt{m^2+p_\t^2}}\left [ x_1 M_0^2-
     p_\t^2+\frac{2m}{w_V}p_\t^2 \right ]\phi_V(x_2,p_{\bot}).
 \eeq
Eq. (\ref{eq:dcV}) coincides with the result in \cite{Jaus2} when $m_1=m_2$. Note that the $1/w_V$
part of Eq. (\ref{eq:dcV}) is different from that in the conventional approach, for example,
\cite{CCH1}. The reason is that the conventional approach is not covariant and contains a spurious
dependence on the orientation of the light front. The relevant calculations are not free of spurious
contributions for transitions involving vector meson. Zero modes, which relate to the $p^-$
integration for $p^+=0$, are required to eliminate the spurious dependence and contribute to the
$1/w_V$ part of Eq. (\ref{eq:dcV}). More detailed discussions about this point can be found in
\cite{Jaus2, CCH2}. The decay constant $f_V$ is related to the electromagnetic decay of vector meson
$V\to e^+e^-$ by \cite{NS}
 \beq \label{eq:dcVexp}
 \Gamma(V\to e^+e^-)=\frac{4\pi}{3}\frac{\alpha^2}{M_V}c_V f_V^2.
 \eeq
where $c_V$ is factor related to the electric charge of the quark
that make up the vector meson.

%%%%%%%%%%%%%%%%%%%%%%%%%%%%%
\subsection{$P\to \gamma\gamma$}
%%%%%%%%%%%%%%%%%%%%%%%%%%%%%

Charge conservation requires charge conjugation $C=+1$ state
coupling to two photons. Thus only pseudoscalar meson can transform
into two photons while the vector meson is forbidden. In the process
of $P\to \gamma\gamma$, the final two photons are both on-shell. For
the purpose of illustration, it is useful to consider a more general
process $P\to\gamma\gamma^*$ with one photon off-shell. We introduce
a transition form factor $F_{P\gamma}(q^2)$ arising from the
$P\gamma\gamma^*$ vertex. The $P\to \gamma\gamma$ process is related
to the form factor at $q^2=0$, i.e., $F_{P\gamma}(0)$. The form
factor $F_{P\gamma}(q^2)$ is defined  by
 \beq
 {\cal A}_\mu=-ie^2F_{{P\gamma}}(q^2)\ep_{\mu\nu\rho\sigma}
  P^{\nu}q_1^{\rho}\ep^{\sigma}.
 \eeq
where ${\cal A}_\mu$ is the decay amplitude of the process
$P\to\gamma\gamma^*$ and $q_1(\ep)$ the momentum (polarization) of
the on-shell photon.

%%%%%%%%%%%%%%%%%%%%%%%%%%%%%%%%%%%%%%%%%%%%%%%%%%%%%%%%%%%%%%%%%%%%%%%%%
\begin{figure}[htbp]
\includegraphics*[width=5in]{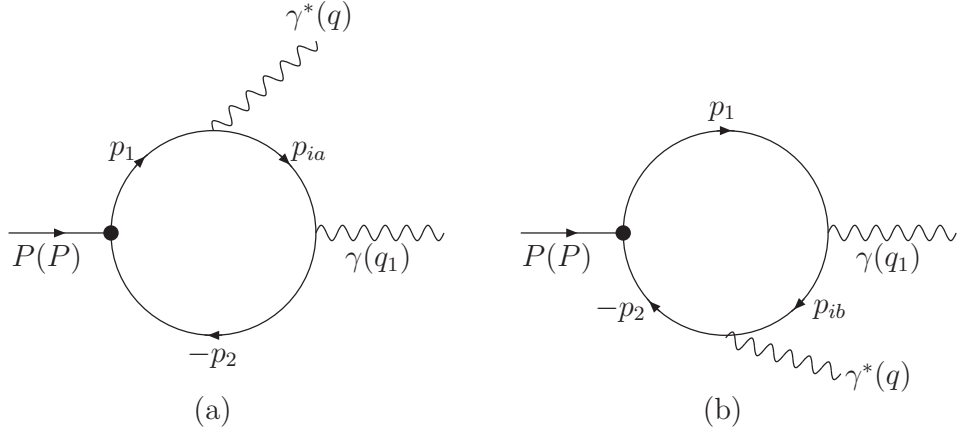}
\caption{Feynman diagram for $P\to\gamma\gamma^*$ process where $P$
in the parenthesis denotes the momentum of meson. The diagram (b) is
related to (a) by the exchange of two photons.}
 \label{fig:Prr}
\end{figure}
%%%%%%%%%%%%%%%%%%%%%%%%%%%%%%%%%%%%%%%%%%%%%%%%%%%%%%%%%%%%%%%%%%%%%%%%%

The transition amplitude for the process of $P\to\gamma\gamma^*$ can
be derived from the common Feynman rules and the vertices for the
meson-quark-antiquark coupling given in Eq. (\ref{eq:HM}). In the
covariant light-front approach, the meson is on-shell while the
constituent quarks are off-shell and the momentum satisfies
$P=p_1+p_2$. To the lowest order approximation, $P\to\gamma\gamma^*$
is a one-loop diagram and depicted in Fig. \ref{fig:Prr}. The
amplitude is given as a momentum integral
 \beq \label{eq:APrr}
  {\cal A}_{\mu}=i e_q^2 e^2 N_c \int \frac{d^4 p_1}{(2\pi)^4}
   \Big \{
   \frac{H_P}{N_1 N_2 N_{ia}} {\rm Tr}[{\gamma_5(-\not \!p_2+m)
   \not \!\ep(\not \!p_{ia}+m)\gamma_{\mu}}(\not \!p_1+m)]  \non \\
     +\frac{H_P}{N_1 N_2 N_{ib}} {\rm Tr}[{\gamma_5(-\not \!p_2+m)
   \gamma_{\mu}} (\not \!p_{ib}+m)\not \!\ep(\not \!p_1+m)]
   \Big \},
 \eeq
where
 \beq
  & p_{ia}=p_1-q,    ~~~~~~~~~~~~~~~~~   & p_{ib}=q-p_2, \non \\
  & N_1=p_1^2-m^2+i\ep,  ~~~~~~~~~~      & N_2=p_2^2-m^2+i\ep, \non \\
  & N_{ia}=p_{ia}^2-m^2+i\ep, ~~~~~~~~~~ & N_{ib}=p_{ib}^2-m^2+i\ep,
 \eeq
and $e_q$ is the electric charge of quark: $e_q=2/3$ for $c$ quark
and $e_q=-1/3$ for $b$ quark. The first and second terms in Eq.
(\ref{eq:APrr}) come from diagrams Fig. \ref{fig:Prr} (a) and (b),
respectively.

For the calculation of the form factor $F_{P\gamma}(q^2)$, it is
convenient to choose the purely transverse frame $q^+=0$, i.e.,
$q^2=-q_{\bot}^2\leq0$. The advantage of this choice is that there
is no the so-called Z-diagram contributions. The price is that only
the form factor at space-like regions can be calculated directly.
The values at the time-like momentum transfer $q^2>0$ regions are
obtained by analytic continuation. In this study, the continuation
is not necessary because we only need the form factors at $q^2=0$
for the $P\to \gamma\gamma$ and $V\to P\gamma$ processes.

At first, we discuss the calculation of Fig. \ref{fig:Prr}(a). The
factors $N_1$, $N_2$ and $N_{ia}$ produce three singularities in the
$p_1^-$ complex plane: one lies in the upper plane; the other two in
the lower plane. By closing the contour in the upper $p_1^-$ complex
plane, the momentum integral can be easily calculated since there is
only one singularity in the plane. This corresponds to putting the
antiquark on the mass-shell. Given this restriction, the momentum
$p_2\to \hat p_2$ with $\hat p_2^2-m^2=0$, and $\hat p_1=P-\hat
p_2$. The on-shell restriction and the requirement of covariance
lead to the following replacements:
 \beq \label{eq:N1}
  N_1 &\to& \hat N_1=x_1(M^2-M_0^2), \non \\
  N_{ia} &\to& \hat N_{ia}=x_2 q^2-x_1 M_0^2+2p_{\bot}\cdot q_{\bot},
    \non \\
  N_2 &\to& \hat N_2=\hat N_1+(1-2 x_1)M^2=x_2 M^2-x_1 M_0^2, \non \\
  \int \frac{d^4 p_1}{(2\pi)^4}\frac{H_P}{N_1N_2N_{ia}} &\to&
  -i\pi\int \frac{dx_2 d^2p_{\bot}}{(2\pi)^4}\frac{h_P}
  {x_2 \hat N_1 \hat N_{ia}}.
 \eeq

For Fig. \ref{fig:Prr}(b), the contour is closed in the lower
$p_1^-$ complex plane. It corresponds to putting the quark on the
mass-shell and the momentum $p_1\to \hat p_1$ with $\hat
p_1^2-m^2=0$. In this case, we need to do the following replacements
 \beq \label{eq:N2}
  N_2 &\to& \hat N_2=x_2(M^2-M_0^2), \non \\
  N_{ib} &\to& \hat N_{ib}=x_1 q^2-x_2 M_0^2
   -2p_{\bot}\cdot q_{\bot}, \non \\
  N_1 &\to& \hat N_1=x_1 M^2-x_2 M_0^2, \non \\
  \int \frac{d^4 p_1}{(2\pi)^4}\frac{H_P}{N_1N_2N_{ib}} &\to&
   -i\pi\int \frac{dx_2 d^2p_{\bot}}{(2\pi)^4}\frac{h_P}
   {x_1 \hat N_2 \hat N_{ib}}.
 \eeq
From Eqs. (\ref{eq:N1}) and  (\ref{eq:N2}), we see that $\hat
N_{ib}$ is obtained from $\hat N_{ia}$ by the exchange of $x_1
\leftrightarrow x_2$ and the change of the sign of $p_{\bot}$.

After the above treatments, the transition amplitude of $P\to
\gamma\gamma^*$ is obtained as
 \beq
 {\cal A}_{\mu}=&&-ie^2\ep_{\mu\nu\rho\sigma}
   P^{\nu}q_1^{\rho}\ep^{\sigma}  \int \frac{dx_2 d^2
   p_{\bot}}{4\pi^3} \frac{N_c e_q^2  m~ h_P}{x_1 x_2 (M^2-M_0^2)}\non\\
   &&~~~~\times\left [ \frac{1}{-x_2 q^2+x_1 M_0^2-2p_{\bot}\cdot q_{\bot}}
    +\frac{1}{-x_1 q^2+x_2 M_0^2+2p_{\bot}\cdot q_{\bot}}
   \right ],
 \eeq

Thus, the final formulae for the form factor $F_{P\gamma}(q^2)$ is
 \beq \label{eq:Pr2}
 F_{P\gamma}(q^2)=&&\frac{e_q^2\sqrt{ 2N_c}}{8\pi^3}\int dx_2 d^2p_{\bot}
   \phi_P(x_2,p_{\bot})\frac{m}{\sqrt{m^2+p_{\bot}^2}} \non\\
   &&~~~~\times\left [ \frac{1}{x_1 M_0^2-x_2 q^2-2p_{\bot}\cdot q_{\bot}}
    +\frac{1}{x_2 M_0^2-x_1 q^2+2p_{\bot}\cdot q_{\bot}}
   \right ],
 \eeq
and $F_{P\gamma}(0)$ is
 \beq \label{eq:Pr20}
  F_{P\gamma}(0)=\frac{e_q^2\sqrt{ 2N_c}}{8\pi^3}\int dx_2 d^2p_{\bot}
   \phi_P(x_2,p_{\bot})\frac{m}{\sqrt{m^2+p_{\bot}^2}}\frac{1}{x_1 x_2 M_0^2}.
 \eeq
By comparing Eq. (\ref{eq:Pr20}) with Eq. (\ref{eq:dcP}), they share
some similarities except the propagators $N_{ia(b)}$ (and a trivial
factor $e_q^2$). This point will become clearer when we discuss the
relation between the light-front QCD and pQCD method.

The decay rate for $P\to \gamma\gamma$ is obtained from the
transition form factors by
 \beq \label{Ptogg}
 \Gamma(P\to \gamma\gamma)=\frac{M_P^3}{64
  \pi}(4\pi\alpha)^2|F_{P\gamma}(0)|^2.
 \eeq

%%%%%%%%%%%%%%%%%%%%%%%%%%%%%
\subsection{$V\to P\gamma$}
%%%%%%%%%%%%%%%%%%%%%%%%%%%%%

Similar to the analysis of $P\to \gamma\gamma$, we also consider a
more general process of $V\to P\gamma^*$ where the final photon is
off-shell. The $V\to P\gamma^*$ transition is parameterized in term
of a vector current form factor $V(q^2)$ by
 \beq
 \Gamma_{\mu}=ie\ep_{\mu\nu\alpha\beta}\ep^{\nu}q^{\alpha}P^{\beta}V(q^2).
 \eeq
where $\Gamma_{\mu}$ is the amplitude of $V\to P\gamma^*$ process.
$P$ ($\ep$) is the momentum (polarization vector) of the initial
vector meson, $P'$ denotes the momentum of the final pseudoscalar
meson, and the momentum transfer $q=P-P'$.
%%%%%%%%%%%%%%%%%%%%%%%%%%%%%%%%%%%%%%%%%%%%%%%%%%%%%%%%%%%%%%%%%%%%%%%%%
\begin{figure}[htbp]
\includegraphics*[width=5in]{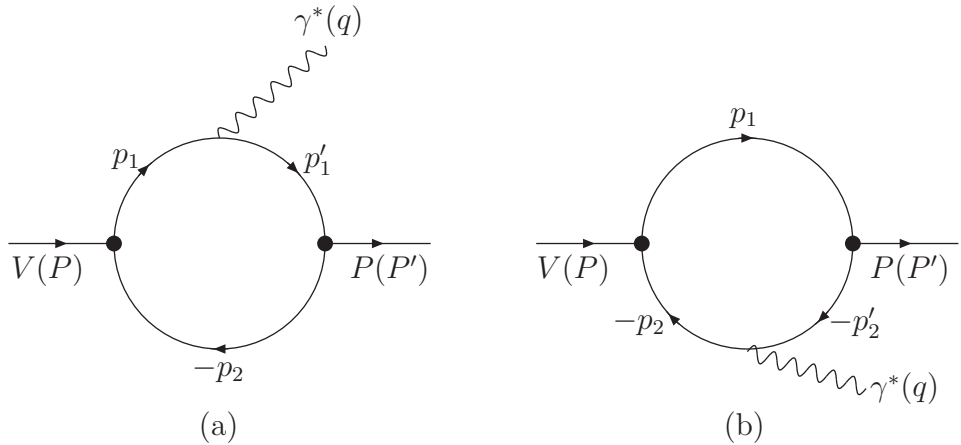}
\caption{Feynman diagram for $V\to P\gamma^*$ process where $P$ in
the parenthesis denotes the momentum of initial meson and $P'$
denotes the momentum of final meson. }
 \label{fig:VPr}
\end{figure}
%%%%%%%%%%%%%%%%%%%%%%%%%%%%%%%%%%%%%%%%%%%%%%%%%%%%%%%%%%%%%%%%%%%%%%%%%
To the lowest order approximation, the $V\to P\gamma^*$ transition
is depicted in Fig. \ref{fig:VPr}. The amplitude $\Gamma_{\mu}$ is
given by a one-loop momentum integral
 \beq \label{eq:AVPr}
 \Gamma_{\mu}=-iee_qN_c \int\frac{d^4 p_1}{(2\pi)^4}
   \Bigg\{ \frac{H_VH_P'}{N_1N_2N_1'}S^a_{\mu\nu}+
           \frac{H_VH_P'}{N_1N_2N_2'}S^b_{\mu\nu}
   \Bigg\} \ep^{\nu},
 \eeq
where
 \beq
 S^a_{\mu\nu}&=&{\rm Tr}
   \Big[ \Big(\gamma_{\nu}-\frac{1}{W_V}(p_1-p_2)_{\nu}\Big)
         (-\not\! p_2+m)\gamma_5(\not \! p_1^{~\prime}+m)
         \gamma_{\mu}(\not \! p_1+m)
   \Big ], \non\\
 S^b_{\mu\nu}&=&{\rm Tr}
   \Big[ \Big(\gamma_{\nu}-\frac{1}{W_V}(p_1-p_2)_{\nu}\Big)
         (-\not \! p_2+m)\gamma_{\mu}(-\not \! p_2^{~\prime}+m)
         \gamma_5(\not \! p_1+m)
   \Big ],
 \eeq
and
 \beq
 N_1'=p_1'^2-m^2+i\ep;    \qquad  \qquad  N_2'=p_2'^2-m^2+i\ep.
 \eeq
The first and second terms in Eq. (\ref{eq:AVPr}) are arising from
diagram (a) and (b) of Fig. \ref{fig:VPr}, respectively. We have
used the momentum relations: $P=p_1+p_2$, $P'=p^{\prime}_1+p_2'$,
$q=P-P'$; $p_2=p_2'$ for diagram Fig. \ref{fig:VPr}(a); $p_1=p_1'$
for diagram Fig. \ref{fig:VPr}(b). It is easy to find that
$S^a_{\mu\nu}=S^b_{\mu\nu}$.

The momentum integral of Eq. (\ref{eq:AVPr}) are performed analogous
to the case of $P\to \gamma\gamma^*$. The contour integrals are
closed in the upper $p_1^-$ half-plane for the first term in Eq.
(\ref{eq:AVPr}) which corresponds to putting antiquark on the
mass-shell; and in the lower half-plane for the second term which
corresponds to putting quark on the mass-shell. For the first term,
it leads to the replacements
 \beq
  N_1^{(\prime)} &\to& \hat N_1^{(\prime)}=x_1 \big(
   M^{(\prime)2}-M_0^{(\prime)2} \big), \non \\
  \int \frac{d^4 p_1}{(2\pi)^4}\frac{H_VH_P'}{N_1N_2N_1'} &\to&
    -i\pi\int \frac{dx_2 d^2p_{\bot}}{(2\pi)^4}\frac{h_Vh_P'}
    {x_2 \hat N_1 \hat N_1'}
 \eeq
In order to preserve the covariance of the decay amplitude, we also
need the replacements
 \beq
 p_1^{\alpha}&\to& x_1 P^{\alpha}-q^{\alpha}
  \frac{p_{\bot}\cdot q_{\bot}}{q^2}, \qquad \qquad
 p_1^{\alpha}p_1^{\beta} \to -g^{\alpha\beta}\left( p_{\bot}^2+
  \frac{(p_{\bot}\cdot q_{\bot})^2}{q^2}\right).
 \eeq
The similar treatments can be done for the second term. After using
the above replacements and Eq. (\ref{eq:htowf}), we obtain the
formulae for the form factor $V(q^2)$ as
 \beq \label{eq:Pr2}
 V(q^2)=&&\frac{e_q}{8\pi^3}\int dx_2 d^2p_{\bot}
   \frac{\phi_V(x_2,p_\bot) \phi'_P(x_2,p'_\bot)}{x_1x_2M_0M_0^{\prime}}
   \left \{ m-\frac{2}{w_V}\left( p_{\bot}^2+
    \frac{(p_{\bot}\cdot q_{\bot})^2}{q^2} \right ) \right \}.
 \eeq
The rate for $V\to P\gamma$ is
 \beq \label{eq:VPr2}
 \Gamma(V\to P\gamma)=\frac{1}{3}\frac{(M_V^2-M_P^2)^3}{32
  \pi M_V^3}(4\pi\alpha)|V(0)|^2.
 \eeq

%%%%%%%%%%%%%%%%%%%%%%%%%%%%%%%%%%%%%%%%%%%%%%%%%%%%%%%%%%
\section{ non-relativistic approximation and perturbative QCD}
%%%%%%%%%%%%%%%%%%%%%%%%%%%%%%%%%%%%%%%%%%%%%%%%%%%%%%%%%%

It is well-known that the system of the heavy quarkonium can be
treated non-relativistically \cite{QR}. A relativistic invariant
theory, light-front QCD in our case, should reproduce the previous
results in the non-relativistic approximations. Here, we will
explore the non-relativistic approximations of the light-front QCD.
It is similar to the studies of heavy quark limit for heavy meson
\cite{CCH1} within the light-front approach. In addition to it, the
light-front QCD is related to perturbative QCD at the large momentum
transfers, such as in $P\to \gamma\gamma$ process. Both of them show
the different aspects of light-front QCD.

At first, we discuss the non-relativistic approximations of the
light-front approach. In the rest frame of the heavy quarkonium, the
momenta of quark and antiquark are dominated by their rest mass
$m\gg \lqcd$ ($\lqcd$ is the hadronic scale). The momentum fractions
$x_1, x_2$ are peaked around $\frac{1}{2}$, and $x_2-\frac{1}{2}$ is
of order $\lqcd/m$. In NRQCD, the velocity of heavy quark is chosen
as the expansion parameter. Neglecting the terms suppressed by
$1/m$, the invariant mass $M_0$ and $p_z$ can be approximated as
 \beq \label{eq:MA}
  M_0\cong 2m\cong M, \qquad \qquad p_z=(x_2-\frac{1}{2})M_0\sim \lqcd.
 \eeq
Compared with $m$, we have neglected the transverse momentum
$p_{\bot}$ because it is of the order of $\lqcd$.  Thus, the
magnitude of the relative momentum $\vec p$ will be much smaller
than $m$, i.e., $|\vec p|\sim \lqcd$, which constitutes the basis of
the non-relativistic treatment.

Under the non-relativistic approximations, the dependence of the
hadron wave function on $x_2$ is replaced by its dependence on $p_z$
since $M_0\cong M$ is a constant. In this way, the hadron wave
function will depend on the relative momentum $\vec p$ only, in
other words, it can be represented by $\psi(\vec p)$. The relation
between the non-relativistic function $\psi(\vec p)$ and the
relativistic one $\phi(x_2, p_{\bot})$ can be established as
follows. From Eq. (\ref{eq:Mpz}), we obtain
 \beq \label{eq:app1}
 dp_z\cong M~ dx_2, \qquad \qquad
 d^3 p=M~ dx_2 d^2p_{\bot},
 \eeq
As uausl, the function of $\psi(\vec p)$ is normalized as
 \beq \label{eq:NRnormal}
 \int \frac{d^3 p}{(2\pi)^3}|\psi(\vec p)|^2=1.
 \eeq
Comparing Eqs. (\ref{eq:app1}) and (\ref{eq:NRnormal}) with Eq.
(\ref{eq:Norm1}), it is straightforward to derive a relation
 \beq
 \phi(x_2, p_{\bot})\doteq \sqrt{2M}\psi(\vec p).
 \eeq
Note that the above relation is valid within the non-relativistic
approximation and is not correct in the general case.

The hadron wave function in the coordinate space $\Psi(\vec r)$ is
obtained by using the Fourier transformation
 \beq
 \Psi(\vec r)=\int \frac{d^3 p}{(2\pi)^3}~\psi(\vec p)~
  e^{i\vec p\cdot \vec r}.
 \eeq
At the origin $\vec r=0$, $\Psi(0)=\int \frac{d^3
p}{(2\pi)^3}~\psi(\vec p)$ is an important parameter which gives the
magnitude of quark-antiquark coupling to the quarkonium. In the
non-relativistic approximations, one can safely neglect $\vec p$
compared to $m$. For example,
 \beq \label{delta}
 \sqrt{m^2+p_{\bot}^2}\to m,~~~ x_2-\frac{1}{2} \to 0.
 \eeq
After these approximations, we can rewrite the decay constants Eqs.
(\ref{eq:dcP}) and (\ref{eq:dcV}) as
 \beq \label{ffPV}
 f_P \doteq 2\sqrt{N_c}\frac{\Psi_P(0)}{\sqrt{M_P}}, \qquad \qquad
 f_V \doteq 2\sqrt{N_c}\frac{\Psi_V(0)}{\sqrt{M_V}}.
 \eeq
Thus,
 \beq
 \frac{f_P^2}{f_V^2}=\frac{M_V}{M_P}\frac{|\Psi_P(0)|^2}{|\Psi_V(0)|^2}.
 \eeq
This is just the so-called Van Royen-Weisskopf formula \cite{RW}.
Since the differences between vector and pseudoscalar vectors arise
from the higher order in $1/m$, these differences vanish in the
limit $m\to \infty$, thus $M_V=M_P=2 m$, $\Psi_V(\vec p)=\Psi_P(\vec
p)$. The ratio of decay constants is equal to 1 in the limit.

For the form factor $V(0)$ of $V\to P\gamma$ process, Eq.
(\ref{eq:Pr2})can be reduced to
 \beq
 V(0)\doteq e_q\int \frac{d^3\vec p}{(2\pi)^3}\frac{2\sqrt{M_P M_V}}{M_V}
      \frac{\Psi_V(\vec p)\Psi_P(\vec p)}{m},
 \eeq
Similarly, in the non-relativistic limit $m\to \infty$, the form
factor $V(0)$ can be further written in a simple form as
 \beq \label{eq:v0}
 V(0)=2e_q/m.
 \eeq
Thus $V(0)$ is a constant, independent of $\Psi(0)$ because of the
normalization condition of $\Psi(\vec p)$. The physical picture is:
the heavy quark and antiquark in the initial and final quarkonium
are in the same momentum configuration at $q^2=0$ point. It is
analogous to the meson system with a single heavy quark that the
Isgur-Wise function is normalized to 1 at the zero-recoil point in
the infinite heavy quark mass limit. From Eqs. (\ref{eq:VPr2}) and
(\ref{eq:v0}), the rate for the $V\to P\gamma$ process is reduced
into
 \beq \label{eq:lVPr}
 \Gamma(V\to P\gamma)=\frac{16}{3}\alpha e_q^2\frac{k_\gamma^3}{M_V^2}.
 \eeq
where $k_\gamma=(M_V^2-M_P^2)/2 M_V$ is the energy of the photon.
This is the leading order result of Eq. (37) in \cite{BJV}.

Next, we discuss that the pQCD is applicable in $P\to \gamma\gamma$
process. In the rest frame of the heavy quarkonium, the total energy
is $2m\gg \lqcd$. Each final photon contains high energy of $m$ and
moves in the opposite light-front direction. When the high energy
photon hits on one nearly rest constituent of the quarkonium, it
causes a large virtuality of the order of $m^2$. In particular, the
virtuality of the internal quark is about $2 m^2$ from Eqs.
(\ref{eq:N1}) and (\ref{eq:N2}). The transverse momentum in the
propagator of the virtual quark can be neglected. Up to leading
order in $\lqcd/m$, the transition form factor $F_{P\gamma}(0)$ is
represented by
 \beq
 F_{P\gamma}(0)&=&e_q^2\sqrt{2N_c}\int\frac{dx_2d^2p_{\bot}}{(2\pi)^3}
   \phi(x_2,p_{\bot})~T_H(x_2) \non\\
 &\propto& \int dx_2~ \Phi(x_2) T_H(x_2).
 \eeq
where $\Phi(x_2)$ is the hadron distribution amplitude obtained from
wave function by integral over the transverse momentum, and
$T_H(x_2)$ is the hard scattering kernel from the subprocess of
$q\bar q\to \gamma\gamma$.

The hard scattering kernel depends on momentum fraction $x_2$ when
the loop corrections are taken into account. But, at tree level, the
hard scattering kernel is
 \beq
  T_H=\frac{1}{m^2},
 \eeq
It is not only independent of transverse momentum $p_{\bot}$ but
also of longitudinal fraction $x_2$. We thus have a further result
 \beq \label{Ftof}
 F_{P\gamma}(0)=e_q^2\frac{f_P}{m^2}.
 \eeq
This equation means that the form factor $F_{P\gamma}(0)$ is
proportional to the decay constant $f_P$ in leading order $\lqcd/m$
and leading order of strong coupling constant $\as$. After combining
Eqs. (\ref{eq:dcVexp}), (\ref{Ptogg}), (\ref{ffPV}), and
(\ref{Ftof}), we finally obtain the decay rates for processes of
$V\to e^+e^-$ and $P\to \gamma\gamma$ as
 \beq
 \Gamma(V\to e^+e^-) &=& {16\over{3}}N_c\pi\alpha^2 c_V \frac{|\Psi_V(0)|^2}{M_V^2},
 \non \\
 \Gamma(P\to \gamma\gamma) &=& 16N_c\pi\alpha^2e_q^4
   \frac{|\Psi_P(0)|^2}{M_P^2}.
 \eeq
These results are the same as ones in Table III in the
non-relativistic quark-potential model \cite{KMRR}.

%%%%%%%%%%%%%%%%%%%%%%%%%%%%%
\section{Numerical results and discussions}
%%%%%%%%%%%%%%%%%%%%%%%%%%%%%

In order to obtain the numerical results, the crucial thing is to
determine the momentum distribution amplitude $\phi(x_2,p_{\bot})$.
One wave function that has been often used in the literature for
mesons is the Gaussian-type
 \beq
 \phi(x_2,p_{\bot})=N\sqrt{\frac{dp_z}{dx_2}}~{\rm exp}\left(
  -\frac{p_{\bot}^2+p_z^2}{2\beta^2} \right),
 \eeq
with $N=4(\pi / {\beta^2})^{3/4}$ and
 \beq
 \frac{dp_z}{dx_2}=\frac{e^2}{x_1x_2 M_0}.
 \eeq
The required input parameters include quark mass: $m_c$ for c quark and $m_b$ for b quark; a hadronic
scale parameter $\beta$ for $\eta_{c(b)}$ and $J/\psi(\Upsilon)$. The quark mass entered into our
analysis is the constituent mass. For light quarks (u and d), the constituent mass which several
hundred MeV, is quite bigger than the current one which is only several MeV obtained from the chiral
perturbation theory. While for the heavy quarks, the difference between them is small. From PDG
\cite{PDG06}, the current masses are $1~{\rm GeV}\leq m_c\leq 1.4~{\rm GeV}$ and $4~{\rm GeV}\leq
m_b\leq 4.5~{\rm GeV}$ in the ${\rm \overline{MS}}$ renormalization scheme. For our purpose, we will
choose  heavy quark constituent masses as
 \beq
  m_c=1.2 ~\GeV,  \qquad \qquad  m_b=4.3 ~\GeV.
 \eeq
Our choices are smaller than the parameters given in \cite{CCH2},
but they are consistent within the error of one $\lqcd$. For the
meson mass, $M_{\eta_c}=2.980~\GeV$, $M_{J/\psi}=3.097~\GeV$ and
$M_{\Upsilon}=9.460~\GeV$ \cite{PDG06}. The mass of $\eta_b$ is
still unknown and it is parameterized as $\Delta
m=M_{\Upsilon}-M_{\eta_b}$. From the references in \cite{ALEPH}, the
range of $\Delta m$ is $\Delta m=30-150$ MeV.

After fixing the quark and meson masses, the remained thing is to
determine the parameters $\beta$. For the vector meson, $\beta_V$ is
extracted from the decay constant $f_V$ which is obtained directly
from the process $V\to e^+e^-$ by Eq. (\ref{eq:dcVexp}). For the
pseudoscalar meson $\eta_c$, $\beta_{\eta_c}$ is extracted from the
decay constant $f_{\eta_c}$ which is obtained from the process $B
\to \eta_c K$.

For the $c\bar c$ charmonium system, there are some experiment data which provides a place to test
the applicability of the Gaussian-type wave function to the heavy quarkonium. From $J/\psi\to
e^+e^-$, we obtain $f_{J/\psi}=416\pm 6$ MeV, and extract $\beta_{J/\psi}=0.639\pm 0.006$ GeV. From
$B \to \eta_c K$, one obtains $f_{\eta_c}=335\pm 75$ MeV \cite{fetac}, and we extract
$\beta_{\eta_c}=0.652^{+0.165}_{-0.143}$ GeV. It is apparent that the dominant errors in the
following calculations will be derived from the uncertainty of $f_{\eta_c}$. By using the above
parameters, we give the numerical results for $\eta_c \to \gamma\gamma$: $Br(\eta_c \to
\gamma\gamma)=(1.78\sim 3.05)\times 10^{-4}$ and for $J/\psi \to \eta_c \gamma$: $Br(J/\psi \to
\eta_c \gamma)=(2.38\sim 2.84)\times 10^{-2}$. The experimental data are $Br(\eta_c \to
\gamma\gamma)=(2.8\pm 0.9)\times 10^{-4}$ and $Br(J/\psi \to \eta_c \gamma)=(1.3\pm 0.4)\times
10^{-2}$. Obviously the former fits experiment very well but the latter does not. This inconsistency
still exists even we adjust the quark mass $m_c$ in the range $1\sim 1.4~\GeV$.

\begin{figure}[htbp]
\includegraphics*[width=4.5in]{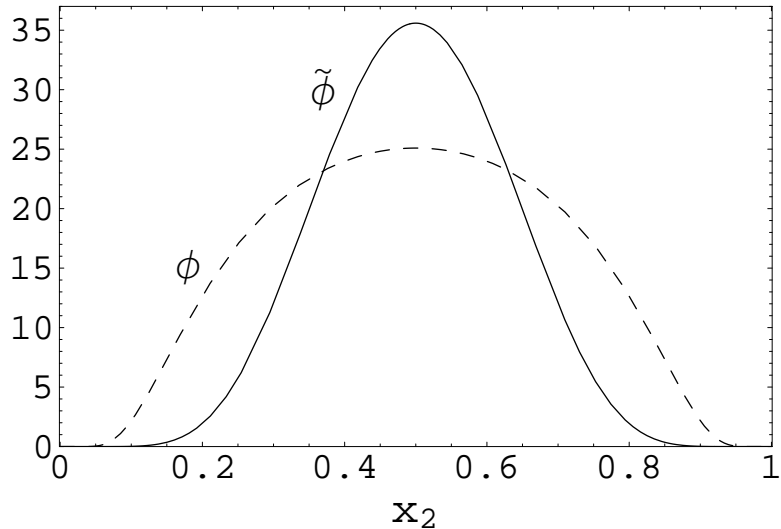}
\caption{The $x$-dependent behaviors of $\phi$ (dash line) and
 $\tilde{\phi}$~(solid line, n=5) at $p^2_\perp=0.1$ GeV$^2$. }
 \label{fig:x-dep}
\end{figure}

One may consider a power law wave function similar to the one employed in Ref. \cite{BS} to fit the
data, however, the Gaussian-type wave function has been used widely in the phenomenal analyses which
related to meson. Thus we modify the Gaussian-type wave function by just multiplying a factor $(x_1
x_2)^n$
 \beq
 \tilde{\phi}(x_2,p_\perp)=\tilde{N}(x_1 x_2)^n \sqrt{\frac{dp_z}{dx_2}}
  ~{\rm exp}\left(-\frac{p_{\bot}^2+p_z^2}{2\tilde{\beta}^2} \right).
 \eeq
The curve which $x_{2}$ is peaked around ${1\over{2}}$ will be
sharped or dulled if $n>0$ or $n<0$, respectively. In the
non-relativistic limit, Eq. (\ref{delta}) reveals that the curve is
near to a delta function $\delta (x_2-{1\over{2}})$. Therefore the
case of $n>0$ seems suitable for the heavy quarkonium. In fact, if
$n=5$ and $m_c=1.2$ GeV, we can extract
$\tilde{\beta}_{J/\psi}=0.786\pm 0.008$ GeV and
$\tilde{\beta}_{\eta_c}=0.807^{+0.273}_{-0.211}$ GeV. The numerical
results $Br(\eta_c \to \gamma\gamma)=(1.56\sim 2.06)\times 10^{-4}$
and $Br(J/\psi \to \eta_c \gamma)=(1.62\sim 2.41)\times 10^{-2}$ are
both consistent with the experimental data. Thus there is a
deduction that, for heavy quarkonium, the momentum fraction $x_2$ is
more centered on ${1\over{2}}$ than one is in the Gaussian-type wave
function. We show the $x$-dependent behaviors of these two types of
wave functions in Fig. \ref{fig:x-dep} and the numerical results in
Table \ref{tab:data-c}.
%%%%%%%%%%%%%%%%%%%%%%%%%%%%%%%%%%%%%%%%%%%%%%%%%%%%%%%%%%%%%%%%%%%%%%%%%
%%%%%%%%%%%%%%%%%%%%%%%%%%%%%%%%%%%%%%%%%%%%%%%%%%%%%%%%%%%%%%%%%%%%%%%%%
%%%%%%%%%%%%%%%%%%%%%%%%%%%%%%%%%%%%%%%%%%%%%%%%%%%%%%%%%%%
\begin{table}[h!]
\caption{\label{tab:data-c} The comparisons between the experimental
data and theory predictions for charmonium decays.}
\begin{ruledtabular}
\begin{tabular}{ccc}
          & $Br(\eta_c \to \gamma\gamma)$
          & $Br(J/\psi\to \eta_c\gamma)$
          \\ \hline
experiment data
          & $(2.8\pm 0.9)\times 10^{-4}$
          & $(1.3\pm 0.4)\%$
          \\
this work $(\phi)$
          & $(1.78\sim3.05)\times 10^{-4}$
          & $(2.38\sim2.84)\%$
          \\
this work $(\tilde{\phi})$
          & $(1.56\sim 2.06)\times 10^{-4}$
          & $(1.62\sim2.41)\%$
\end{tabular}
\end{ruledtabular}
\end{table}
%%%%%%%%%%%%%%%%%%%%%%%%%%%%%%%%%%%%%%%%%%%%%%%%%%%%%%%%%%%%%%

\begin{figure}[htbp]
\includegraphics*[width=4.5in]{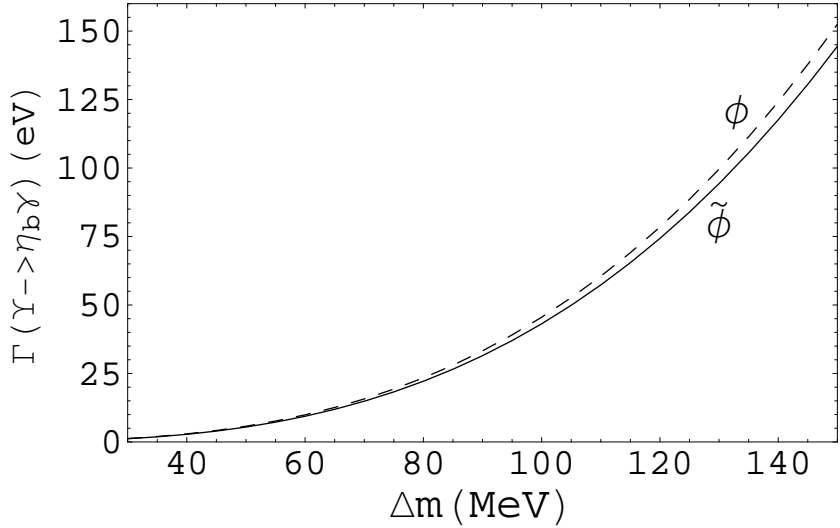}
\caption{ The dependences of $\Gamma(\Upsilon\to \eta_b\gamma)$ on $\Delta
m=M_{\Upsilon}-M_{\eta_b}$.}
 \label{fig:VP-dm}
\end{figure}

For the $b\bar b$ bottomonium system, the experimental data are relatively less. From
$\Upsilon(1S)\to e^+e^-$, we obtain $f_{\Upsilon}=708\pm 8$ MeV, then extract
$\beta_{\Upsilon}=1.323\pm 0.010$ GeV and $\tilde{\beta}_{\Upsilon}=1.463\pm 0.012$ GeV. However,
$\eta_b$ meson hasn't been observed in experiment. It is impossible to determine the decay constant
$f_{\eta_b}$ from the experiment. As had been discussed, the relation $f_{\eta_b}=f_{\Upsilon}$ is
hold in the non-relativistic limit. Since the corrections to this relation are suppressed by
$\lqcd/m_b$, it may be reasonable to use it to determine the parameter
$\beta_{\eta_b}(\tilde{\beta}_{\eta_b})$. Therefore we obtain $\beta_{\eta_b}=1.433\pm 0.014$ and
$\tilde{\beta}_{\eta_b}=1.607\pm 0.018$ GeV. For obtaining the decay widths $\Gamma(\eta_b \to
\gamma\gamma)$ and $\Gamma(\Upsilon\to \eta_b\gamma)$, we must be aware of the value of $\Delta m$.
However, the sensitivities of these two decay widths to $\Delta m$ are quite different. On the one
hand, $\Gamma(\eta_b \to \gamma\gamma)$ is insensitive to $\Delta m$ because $M_{\eta_b} \gg \Delta
m$ (see Eq. (\ref{Ptogg})). On the other hand, $\Gamma(\Upsilon\to \eta_b\gamma)$ is very sensitive
to $\Delta m$ because it is proportional to $(\Delta m)^3$ (see Eq. (\ref{eq:VPr2})). Thus here we
list the values of $\Gamma(\eta_b \to \gamma\gamma)$ for $\Delta m=0.09\pm 0.06$ GeV and
$\Gamma(\Upsilon\to \eta_b\gamma)$ for $\Delta m=0.09$ GeV in Table \ref{tab:data}. The dependences
of $\Upsilon\to \eta_b\gamma$ on $\Delta m$ are also shown in Fig. \ref{fig:VP-dm}.

%%%%%%%%%%%%%%%%%%%%%%%%%%%%%%%%%%%%%%%%%%%%%%%%%%%%%%%%%%%
\begin{table}[h!]
\caption{\label{tab:data} The comparisons among the several theory
predictions for bottomonium decays.}
\begin{ruledtabular}
\begin{tabular}{ccc}
          & $\Gamma(\eta_b \to \gamma\gamma) $ (eV)
          & $\Gamma(\Upsilon\to \eta_b\gamma)$ (eV)
          \\ \hline
this work $(\phi)$
          & $453 \pm 17$
          & $33.2 \pm 0.1$
          \\
this work $(\tilde{\phi})$
          & $422 \pm 15$
          & $31.5 \pm 0.1$
          \\
used in \cite{ALEPH}
          & $557\pm 85$
          & -
          \\
NRQCD \cite{NRQCD1}$\cal{O}(\alpha_{\mathrm s})$
          & $460$
          & -
          \\
potential model \cite{potential}
          & $466\pm 101$
          & -
          \\
\end{tabular}
\end{ruledtabular}
\end{table}
%%%%%%%%%%%%%%%%%%%%%%%%%%%%%%%%%%%%%%%%%%%%%%%%%%%%%%%%%%%%%%
%%%%%%%%%%%%%%%%%%%%%%%%%%%%%%%%%%%%%%%%%%%%%%%%%%%%%%%%%%%%%%

For the numerical results, some comments are in orders:

(1) The decay constant for $\eta_c$ is $f_{\eta_c}=335\pm 75$ MeV
\cite{fetac}, and we obtain
 \beq
 \left(\frac{f_{\eta_c}}{f_{J/\psi}}\right)^2\approx 0.65\pm 0.31.
 \eeq
%This result is consistent with $1.5\pm 0.1$ in \cite{AM} and bigger
%than $1.06\pm 0.14$ in \cite{HK}.
The difference between the pseudoscalar and vector meson in
light-front approach comes from power suppressed terms: the
transverse momentum $p_\t$, $x_2-\frac{1}{2}$ and the wave function
$\phi(x,p_\t)$. The deviation of the results from 1 shows that
$\lqcd/m_c\sim 30\%$ corrections cannot be neglected.
%For $b\bar b$ system, $(f_{\eta_b}/f_{\Upsilon})^2=1$ is our assumption and
%$1.16\pm 0.06$ in \cite{AM} and $0.99\pm 0.04$ \cite{HK}.

(2) For the M1 transition $J/\psi\to \eta_c\gamma$, the leading order prediction from Eq.
(\ref{eq:lVPr}) for the branching ratio is $4.2\%$ which is about a factor of 3 larger than the
experimental data. This means that the next-to-leading order $\lqcd/m_c$ corrections are so
substantial that they must be included in the calculations.

(3) For the decay width $\Gamma(\eta_b \to \gamma\gamma)$, it is
insensitive to the variations of $\Delta m$ but proportional to
$f_{\eta_b}^2$. Thus an adjustment of $f_{\eta_b}$ by $10\%$ will
correspond to a variation of the theory prediction by about $20\%$.
So far, $(f_{\eta_b}/f_{\Upsilon})^2=1$ is our assumption and
$0.99\pm 0.04$ in \cite{HK} and $1.16\pm 0.06$ in \cite{AM}. After
considering these uncertainties, our prediction may be consistent
with previous results listed in \cite{ALEPH,NRQCD1,potential}.
%%%%%%%%%%%%%%%%%%%%%%%%%%%%%
\section{Conclusions}
%%%%%%%%%%%%%%%%%%%%%%%%%%%%%

In this article we have studied the decay constants, two-photon annihilation $P\to \gamma\gamma$ and
magnetic dipole transition $V\to P\gamma$ processes for the ground-state heavy quarkonium within the
covariant light-front approach. The phenomenological parameters and wave functions are determined
from the experiment. The predictions agree with the measured data within the theoretical and
experimental errors. The quark mass we use is very close to the current mass which is different from
the choices in non-relativistic quark model. The difference between the $J/\psi$ and $\eta_c$ decay
constants and the study in $J/\psi\to \eta_c\gamma$ both show that the power corrections from wave
functions and transverse momentum effects are important. In order to make a better fit to the
experimental data, we adjust the longitudinal momentum fraction of the wave function to center around
$1/2$ further. We also give a numerical prediction for $\eta_b\to \gamma\gamma$ and $\Upsilon\to
\eta_b\gamma$. The branching ratio for $\Upsilon\to \eta_b\gamma$ is too small to be observed.
$\eta_b\to \gamma\gamma$ may be a good process to determine $\eta_b$ and its mass. The QCD
corrections are neglected in this study, including them will slightly change the wave function inputs
but does not change our conclusions.

The light-front approach shows different aspects of QCD. Under the
non-relativistic approximations, the light-front approach reproduces
the results in the non-relativistic quark-potential model. For $P\to
\gamma\gamma$ where two final photons are on the opposite
light-front, the process is perturbative dominated and the
light-front approach reduces to the model-independent pQCD. The
light-front method unifies the perturbative and non-perturbative QCD
into the same framework.

We have considered s-wave heavy quarkonium in the light-front approach only, the applications to
other quantities and higher resonances are in progress. One interesting thing may be to explore the
light-front approach in NRQCD (or pNRQCD). This will provide an alternative non-perturbative method
to calculate the hadronic matrix elements defined in NRQCD.

\vspace{0.5cm} {\bf Acknowledgments}\\
We thank Hai-Yang Cheng and Chun-Khiang Chua for many valuable
discussions. We also wish to thank the National Center for
Theoretical Sciences (South) for its hospitality during our summer
visits where this work started. This work was supported in part by
the National Science Council of R.O.C. under Grant No.
NSC94-2112-M-017-004.


\begin{thebibliography}{99}

\bibitem{QR} C. Quigg and J. L. Rosner, Phys. Rept. {\bf 56}, 167-235
 (1979).

\bibitem{HQ} N. Brambilla {\it et al.}, CERN-2005-005, [hep-ph/0412158].

\bibitem{BPP} S. J. Brodsky, H. C. Pauli and S. S. Pinsky, Phys. Rept. {\bf 301}, 299 (1998).

\bibitem{LFQM} M. V. Terent'ev, Sov. J. Phys. {\bf 24}, 106 (1976); V. B.
 Berestetsky and M. V. Terent'ev, {\it ibid.}{\bf 24}, 547 (1976);
 {\it ibid.} {\bf 25}, 347 (1977).

\bibitem{Jaus1} W. Jaus, Phys. Rev. D{\bf 41}, 3394 (1990); {\it ibid.} {\bf 44}, 2851
  (1991).

\bibitem{CCH1} H. Y. Cheng, C. Y. Cheung and C. W. Hwang, Phys. Rev. D {\bf 55}, 1159 (1997).

\bibitem{Jaus2} W. Jaus, \prd {\bf 60}, 054026 (1999).

\bibitem{CCH2} H. Y. Cheng, C. K. Chua and C. W. Hwang, Phys. Rev. D {\bf 69}, 074025 (2004).

\bi{Hwang} C. W. Hwang, \prd{\bf 64}, 034011 (2001).

\bi{KMRR} W. Kwong, P. B. Mackenzie, R. Rosenfeld and J. L. Rosner,
 \prd {\bf 37}, 3210 (1988).

\bi{AB} E. S. Ackleh, T. Barnes, \prd{\bf 45}, 232 (1992).

\bibitem{BJV} N. Brambila, Y. Jia and A. Vairo, \prd {\bf 73}, 054005
 (2006).

\bi{DER} J. J. Dudek, R. G. Edwards and D. G. Richards,
 \prd{\bf 73}, 074507 (2006).

\bibitem{ALEPH} ALEPH Collaboration, \plb {\bf 530} 56-66 (2002).

\bibitem{PB} H. C. Pauli and S. J. Brodsky, Phys. Rev. {\bf D32}, 1993; 2001 (1985).

\bibitem{TBP} A. C. Tang, S. J. Brodsky, and H. C. Pauli, Phys. Rev. {\bf D44}, 1842 (1991).

\bibitem{KP} M. Kaluza and H. C. Pauli, Phys. Rev. {\bf D45}, 2968 (1992).

\bibitem{KPP} M. Kaluza and H.-J. Pirner, Phys. Rev. {\bf D47}, 1620 (1993).

\bibitem{PHW} R. J. Perry, A. Harindranath, and K. G. Wilson, Phys. Rev. Lett. {\bf 65}, 2959 (1990).

\bibitem{PH} R. J. Perry and A. Harindranath, Phys. Rev. {\bf D43}, 4051 (1991).

\bibitem{GHPSW} S. G{\l}azek, A. Harindranath, S. Pinsky, J. Shigemitsu, and K. G. Wilson, Phys. Rev. {\bf D47},
1599 (1993).

\bibitem{T} I. Tamm. J. Phys. (USSR) {\bf 9}, 449 (1949).

\bibitem{D} S. M. Dancoff, Phys. Rev. {\bf 78}, 382 (1950).

\bibitem{EPB} T. Eller, H. C. Pauli, and S. J. Brodsky, Phys. Rev. {\bf D35}, 1493 (1987).

\bi{NS} M. Neubert and B. Stech, Adv. Ser. Direct. High Energy
 Phys. {\bf 15}, 294-344 (1998).

\bibitem{RW} R. Van Royen and V. F. Weisskopf, Nuovo Cim. {\bf 50}, 617
 (1967); $ibid$.{\bf 51}, 583E (1967).

\bibitem{PDG06} Particle Data Group 2006, W.-M. Yao {\it et al.},
 Journal of Physics G {\bf 33}, 1 (2006).

\bibitem{fetac} K. W. Edwards {\it et al.}, CLEO Collaboration,
 \prl {\bf 86}, 30-40 (2001).

\bibitem{BS} S. J. Brodsky and F. Schlumpf, Phys. Lett. B {\bf 329}, 111-116, (1994).

\bibitem{NRQCD1} G. A. Schuler, F. A. Berends, and R. van Gulik,
 Nucl. Phys. B {\bf 523}, 423-438 (1998).

\bibitem{potential} N. Fabiano, Eur. Phys. J. C {\bf 26}, 441-444 (2003).

\bibitem{HK} D. S. Hwang and G. H. Kim, Z. Phys. C {\bf 76}, 107-110 (1997).

\bibitem{AM}  M. R. Ahmady and R. R. Mendel, \prd {\bf 51}, 141 (1995).

\end{thebibliography}
\end{document}